\documentclass[a4paper,12pt]{article}
\pdfoutput=1 % if your are submitting a pdflatex (i.e. if you have
\usepackage{graphicx}
\usepackage{tikz}
\usepackage{jheppub} % for details on the use of the package, please
\newcommand{\be}{\begin{equation}}
\newcommand{\eeq}{\end{equation}}
\newcommand{\bea}{\begin{eqnarray}}
\newcommand{\eea}{\end{eqnarray}}
\newcommand{\ba}{\begin{array}}
\newcommand{\ea}{\end{array}}

\newcommand{\ee}{\end{equation} }

\newcommand{\tr}{\mathrm{tr}\,}

\newcommand{\one}{{\rm 1\kern -.9mm l}}

\title{Deformed SW curve and the null vector decoupling equation
in Toda field theory}

\author{Rubik Poghossian}

\affiliation{Yerevan Physics Institute,\\
Alikhanian Br. 2, AM-0036 Yerevan, Armenia}
\emailAdd{poghos@yerphi.am}

\abstract{It is shown that the deformed Seiberg-Witten curve equation after Fourier transform is
mapped into a differential equation for the AGT dual 2d CFT cnformal block containing an extra completely degenerate field. We carefully match parameters in two sides of duality
thus providing not only a simple independent prove of the AGT correspondence in Nekrasov-Shatashvili limit, but also an extension
of AGT to the case when a secondary field is included in the CFT
conformal block. Implications of our results in the study of
monodromy problems for a large class of $n$'th order Fuchsian
differential equations are discussed.
}

\keywords{Deformed Seiberg-Witten equation, Toda field theory, AGT, Fuchsian differential equations, Accessory parameters}

\preprint{YerPhI/2016/01}
\begin{document}

\maketitle
\flushbottom
\section{Introduction}
Low energy behavior of $N=2$ SYM theory admits an exact description
including both perturbative and non-perturbative layers \cite{Seiberg:1994aj,Seiberg:1994rs}. All relevant
quantities, such as the prepotential and chiral gauge invariant
expectation values are nicely encoded in the geometry of Riemann
surfaces, called in this context the Seiberg-Witten curve. It was
realized from the very beginning that this curve is intimately related
to the classical integrable systems \cite{Gorsky:1995zq,Martinec:1995by}.
Later development of this field
was triggered by the application of localization method
\cite{Nekrasov:2002qd,Flume:2002az,Bruzzo:2002xf,Nekrasov:2003rj}. An earlier important reference is \cite{Lossev:1997bz}. To make localization
method efficient one should first formulate the theory in a non-trivial
background, commonly referred as the $\Omega$-background \cite{Nekrasov:2002qd,Nekrasov:2003rj}, which is
parameterized by two numbers $\epsilon_1$, $\epsilon_2$ (these are
rotation angles in $(x^1,x^2)$ and $x^3,x^4$ planes of Euclidean
space time). The $\Omega $-background brakes the Poincar\'e symmetry
and effectively regularizes the space-time volume, making the
partition function finite. Using localization the instanton part of the
partition function as well as the chiral correlators of the theory
can be represented as sum over arrays of Young diagrams in such a way,
that their total number of boxes coincides with the instanton number.
Sending the parameters $\epsilon_{1,2}$ to zero one recovers the
known results of the trivial background. It appears, nevertheless, that
the theory on finite $\Omega$-background has its own significance.
Namely, the recent developments recovered intriguing relations of these
theory with 2d CFT called the AGT correspondence \cite{Alday:2009aq,Wyllard:2009hg,Fateev:2011hq}. According
to AGT correspondence the partition functions of gauge theories get
identified with 2d CFT conformal blocks.

An interesting special case of the $\Omega$-background is the Nekrasov-
Shatashvili limit \cite{Nekrasov:2009rc} when only one of the parameters, say
$\epsilon_1\rightarrow 0$. In this limit the classical integrable system
associated with SW curve gets quantized so that the remaining parameter
$\epsilon_2$ plays the role of the Plank's constant. In
\cite{Mironov:2009uv,Mironov:2009dv,Maruyoshi:2010iu,Marshakov:2010fx}
this limit has been investigated using Bohr-Sommerfeld semiclassical
method. Another approach initiated in \cite{Poghossian:2010pn}
and further developed in \cite{Fucito:2011pn,Fucito:2012xc,Nekrasov:2013xda}
is based on the careful analysis of the contributions of
various arrays of Young diagrams. It was shown in \cite{Poghossian:2010pn} that there is
a single array of diagrams which dominates in the NS limit. This approach leads to a generalization of the notion of Seiberg-Witten curve. The
algebraic equations defining SW curve get replaced by difference equations
(referred as deformed Seiberg-Witten curve or shortly DSW equations). It is
worth noting that like the original SW curve, DSW "curve" besides the
prepotential encodes information about all chiral correlators. Let me
describe briefly how DSW equation emerges. One starts with an entire
function whose zeros encode the lengths of the rows of the dominant
array of Young diagrams mentioned above. The condition of giving the most
important contribution is translated then to a linear difference equation
for this entire function \cite{Poghossian:2010pn}. The DSW equation (no longer linear)
emerges as a condition on the ratio of this entire function with
itself with a shifted argument. The initial linear difference
equation is closely related to the Baxter's $T-Q$ equation
which plays an important role in the context of exactly integrable
statistical mechanics \cite{Baxter:1982} and QFT models \cite{Bazhanov:1998dq}.
Fourier transform of the linear difference leads to a linear
differential equation. This is the same equation which emerged
as the Schr\"odinger equation in the already mentioned
alternative Bohr-Sommerfeld approach to the NS limit.

For the purposes of this paper it is essential that from the
AGT perpective the NS limit corresponds to the classical
($c\rightarrow \infty $) limit of 2d CFT conformal block of "heavy" fields.
The idea to apply DSW equation to investigate semiclassical limit of
2d CFT was suggested in \cite{Poghossian:2010pn}. For the alternative
approaches to the NS limit and the semiclassical CFT see e.g. \cite{Maruyoshi:2010iu,Marshakov:2010fx,Piatek:2011tp,
Bulycheva:2012ct,Piatek:2013ifa}.
The case of irregular conformal blocks is considered in \cite{Choi:2015idw}.
From the AGT point of view the linear differential equation discussed
in previous paragraph appears to be closely related to the null-vector
decoupling equation in 2d CFT. Some results in this direction has been
already announced in \cite{Nekrasov:2013xda}. For applications
of CFT degenerate fields in AGT context see also
\cite{Bonelli:2011fq,Bonelli:2011wx,
Fucito:2013fba,Bershtein:2014qma,Ashok:2015gfa}.

  In this paper we systematically
Investigate this relationship in a quite general setting of $A_r$ linear
quiver theories with an arbitrary number (equal to $r$) of $SU(n)$ gauge
groups corresponding in AGT dual CFT side to the $r+3$-point conformal
blocks in $W_n$ Toda field theory.

The subsequent material is organized as follows.
In Section \ref{section2} we
investigate DSW equations for $A_r$ quiver theories and establish explicit
relations between curve parameters and chiral expectation values. Then
using Fourier transform we derive the corresponding linear differential
equation and thoroughly investigate its singular points. In
Section \ref{section3} starting from the general structure of the fusion rule of the
completely degenerated field $V_{-b\omega_1}$ and using Ward identities
together with some general requirements necessary to get acceptable
solutions,  we derive the null-vector decoupling equation in the
{\it semiclassical} limit. Note that our approach here is somewhat
heuristic and seems to be applicable only in semiclassical case.
To get exact differential equation valid in full pledged
quantum case one should construct the null vector explicitly and make
use of the complicated $W_n$-algebra commutation relations and Ward
identities. To my knowledge, at least in its full generality, this goal
has not been achieved  yet. In Section \ref{section4} we show that under a simple
transformation the two differential equation of previous chapters can
be completely matched. Already comparison of the first non-trivial
coefficient functions (in 2d CFT side this function is the classical expectation
value of the stress-energy tensor) readily recovers the celebrated AGT correspondence. Thus our analysis provides a new, surprisingly
elementary proof of the AGT duality in semiclassical limit.
Matching further coefficient functions (i.e. the higher spin
$W$-current expectation values in 2d CFT side and their gauge theory
counterparts) extends the scope of AGT correspondence: the conformal blocks including a descendant field get related to the higher power chiral expectation values in gauge theory. This new relations are explicitly demonstrated in full details in the case $r=1$ corresponding to the
four-point conformal blocks. Finally in {\it Conclusion} we emphasize the
relevance of our findings in the context of the monodromy problems in a large class of Fuchsian differential equations.

%%%%%%%%%%%%%%%%%%%%%%%%%%%%%%%%%%%%%%%%%%%%%%%%%%%%%%%%%%%%%%%%%%%%%%%%%%%%%
%%%%%%%%%%%%%%%%%%%%%%%%%%%%%%%%%%%%%%%%%%%%%%%%%%%%%%%%%%%%%%%%%%%%%%%%%%%%%

\section{Deformed Seiberg-Witten curve for $A_r$ quiver}
\label{section2}
Partition function and chiral correlators of ${\cal N}=2$ gauge theory
can be represented as sum over arrays of Young diagrams which label
the fixed points of space time rotations and global
gauge transformations acting in the moduli space of instantons.
\cite{Nekrasov:2002qd,Flume:2002az,Bruzzo:2002xf,Nekrasov:2003rj}.
%%%%%%%%%%%%%%%%%%%%%%%%%%%%%%%%%%%%%%%%%%%%%%%%%%%%%%%%%%%%%%%%%%%%%%%%%%%%%
%%%%%%%%%%%%%%%%%%%%%%%%%%%%%%%%%%%%%%%%%%%%%%%%%%%%%%%%%%%%%%%%%%%%%%%%%%%%%
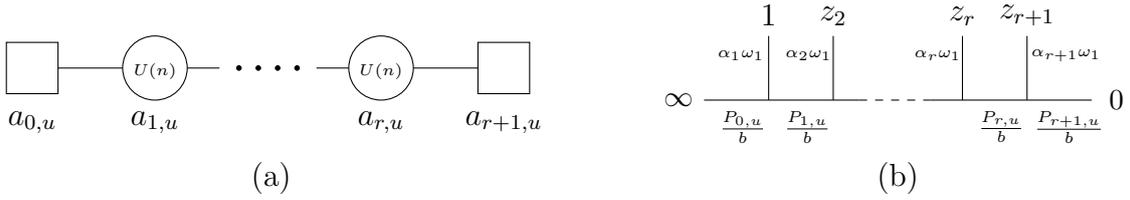
\begin{figure}[t]
\begin{tikzpicture}[scale=0.85]
%\draw[help lines] (0,0) grid (23,1);
% Quiver gauge theory
\draw [] (0.2,0.1) rectangle (1,0.9);
\node [below] at (0.6,0) {$a_{0,u}$};
\draw [] (1,0.5)--(2,0.5);
\draw [] (2.5,0.5) circle [radius=0.5];
\node [] at (2.5,0.45) {\tiny{$U(n)$}};
\node [below] at (2.5,0) {$a_{1,u}$};
\draw [] (3,0.5)--(3.5,0.5);
\draw [fill=black] (3.8,0.5) circle [radius=1pt];
\draw [fill=black] (4.1,0.5) circle [radius=1pt];
\node [below] at (4.3,-0.8) {(\cal{a})};
\draw [fill=black] (4.4,0.5) circle [radius=1pt];
\draw [fill=black] (4.7,0.5) circle [radius=1pt];
\draw [] (5,0.5)--(5.5,0.5);
\draw [] (6,0.5) circle [radius=0.5];
\node [] at (6,0.45) {\tiny{$U(n)$}};
\node [below] at (6,0) {$a_{r,u}$};
\draw [] (6.5,0.5)--(7.5,0.5);
\draw [] (7.5,0.1) rectangle (8.3,0.9);
\node [below] at (7.9,0) {$a_{r+1,u}$};
%CFT block
\draw [] (11,0)--(13.4,0);
\draw [dashed] (13.4,0)--(14.5,0);
\draw [] (14.5,0)--(17,0);
\node [left] at (11,0) {\small{$\infty$}};
\node [below] at (11.6,0) {\scriptsize{$\frac{P_{0,u}}{b}$}};
\node [below] at (12.6,0) {\scriptsize{$\frac{P_{1,u}}{b}$}};
\node [below] at (15.6,0) {\scriptsize{$\frac{P_{r,u}}{b}$}};
\node [below] at (16.65,0) {\scriptsize{$\frac{P_{r+1,u}}{b}$}};
\node [right] at (17.1,0) {\small{$0$}};
\draw [] (12,0)--(12,1);
\node [left] at (12.1,0.7) {\tiny{$\alpha_1\omega_1$}};
\node [above] at (12,1) {$1$};
\draw [] (13,0)--(13,1);
\node [left] at (13.15,0.7) {\tiny{$\alpha_2\omega_1$}};
\node [above] at (13,1) {$z_2$};
\draw [] (15,0)--(15,1);
\node [left] at (15.15,0.7) {\tiny{$\alpha_{r}\omega_1$}};
\node [above] at (15,1) {$z_{r}$};
\draw [] (16,0)--(16,1);
\node [right] at (15.9,0.7) {\tiny{$\alpha_{r+1}\omega_1$}};
\node [above] at (16,1) {$z_{r+1}$};
\node [below] at (14,-0.8) {(\cal{b})};
\end{tikzpicture}
\caption{(\cal{a}) The quiver diagram for the conformal
linear quiver $U(n)$ gauge theory:
$r$ circles stand for gauge multiplets; two squares represent $n$ anti-fundamental (on the left edge) and $n$ fundamental (the right edge) hypermultiplets; the lines connecting adjacent circles are the bi-fundamentals.
(\cal{b}) The AGT dual conformal block of the Toda field theory.}
\label{quiv_block}
\end{figure}
%%%%%%%%%%%%%%%%%%%%%%%%%%%%%%%%%%%%%%%%%%%%%%%%%%%%%%%%%%%%%%%%%%%%%%%%%%%
%%%%%%%%%%%%%%%%%%%%%%%%%%%%%%%%%%%%%%%%%%%%%%%%%%%%%%%%%%%%%%%%%%%%%%%%%%%
In the case of $A_r$ quiver theory with fundamental and bi-fundamental matter hypermultiplets and unitary $U(n)$ gauge groups
(see Fig.\ref{quiv_block}\cal{a}), there is an $n$-tuple of Young diagrams associated
to each of the $r$ gauge groups (indicated by circles in
Fig.\ref{quiv_block}\cal{a}).
It has been shown in \cite{Poghossian:2010pn} for the case of a single
gauge group and later generalized further
in \cite{Fucito:2011pn,Fucito:2012xc,Nekrasov:2013xda}
that among all fixed points in moduli space of
instantons there is a unique one giving a non-vanishing contribution in the
Nekrasov Shatashvili limit\footnote{For simplicity in this
paper we'll set $\epsilon_2=1$. This is not a loss of generality
since a generic $\epsilon_2$ everywhere can be restored
by simple scaling arguments.}. We will denote the (rescaled by
$\epsilon_1$) lengths of the rows of this
``critical'' array of Young diagrams $Y_{\alpha,u}$ by $\lambda_{\alpha,u,i}$ where
$\alpha =1,\ldots r $ refers to the node of the quiver, $u=1,\ldots ,n $
is the index of the defining representation of the gauge group $U(n)$
associated with this node and $i=1,2,\ldots $ specifies the row. The data
$\lambda_{\alpha,u,i}$ can be conveniently encoded in meromorphic functions
$y_\alpha(x)$, which are endowed with zeros located at $x=a_{\alpha,u}+i-1
+\lambda_{\alpha,u,i}$ and poles at $x=a_{\alpha,u}+
i-2+\lambda_{\alpha,u,i}$ where $a_{\alpha,u}$ are the Coulomb branch
parameters. In addition we associate to the "frozen" nodes
(indicated by squares in Fig.\ref{quiv_block}\cal{a}) the parameters $a_{0,u}$ and $a_{r+1,u}$.
In terms of these parameters the masses of fundamental and anti-fundamental
hypermultiplets are given by
\bea
m_u=a_{r+1,u}-\frac{1}{n}\sum_{v=1}^n a_{r,u}\quad \text{and}\quad \bar{m}_u=a_{0,u}-\frac{1}{n}\sum_{v=1}^n a_{1,v}
\eea
respectively. In terms of
\bea
\bar{a}_\alpha=\frac{1}{n}\sum_{u=1}^n a_{\alpha,u}
\eea
the masses of the bifundamental hypermultiplets are simply
\bea
m_{\alpha,\alpha+1}=\bar{a}_{\alpha+1}-\bar{a}_\alpha\,.
\eea
The Deformed Seiberg-Witten (DSW) curve equations arise from the
condition on the  instanton configuration to give the most important contribution to
the prepotential in NS limit. In the case of our present interest
of $A_r$ quiver theory we get a system of $r$ (difference) equations for
$r$ functions $y_\alpha(x)$, $\alpha=1, \ldots ,r$. In addition we introduce two polynomials
\bea
y_0(x)=\prod_{u=1}^n(x-a_{0,u}); \qquad
y_{r+1}(x)=\prod_{u=1}^n(x-a_{r+1,u})
\label{y0_yrp1_factorized}
\eea
which encode fundamental hyper-multiplets attached
to the first and the last nodes of the quiver
Fig.\ref{quiv_block}\cal{a}. The equations can be found using iterative procedure based on so called iWeyl reflections
(i stands for instanton) \cite{Nekrasov:2012xe}
\bea
y_\alpha(x)\rightarrow y_\alpha(x)+\frac{y_{\alpha-1}(x-1)y_{\alpha+1}(x)}
{y_\alpha(x-1)}
\eea
It appears that the result of this procedure is related to the $q$-Character
of the $\alpha$'th fundamental representation of
the group $A_{r}$. Explicitly for $\alpha=1,2,\ldots, r$
one obtains\footnote{When comparing this formula with those of
\cite{Fucito:2012xc,Nekrasov:2013xda} it should be taken into
account that we have shifted arguments in $y_\alpha(x)$, $\chi_\alpha(x)$
appropriately to get rid of explicit appearance of the
bifundamental masses.}
\bea
\chi_\alpha(x)=y_0(x-\alpha)\sum_{1\le k_1<k_2<\cdots <k_{\alpha}\le r+1}
\prod_{\beta=1}^\alpha
\left(\frac{y_{k_\beta}(x-\alpha+\beta)}{y_{k_\beta-1}(x-\alpha+\beta-1)}\,\,\,
q_\beta^{\beta-\alpha}\prod_{\gamma=1}^{k_\beta-1}q_\gamma\right),\quad
\label{chi_alpha}
\eea
where $q_{\alpha}$ are the gauge couplings,
$\chi_\alpha(x)$ are $n$-th order polynomials in $x$ with
coefficients related to the expectation values
$\langle\tr \phi_\alpha^J\rangle $ ($\phi_\alpha$ are the scalars of
the vector multiplet) in a way to be specified below.
For later purposes we'll set by definition  $\chi_{0}(x)\equiv y_{0}(x)$ and
 $\chi_{r+1}(x)\equiv y_{r+1}(x)$. The difference equations (\ref{chi_alpha})
are the deformed Seiberg-Witten equations \cite{Poghossian:2010pn} for the
$A_r$ quiver gauge theory \cite{Fucito:2012xc,Nekrasov:2013xda}.
It is assumed that all functions $y_\alpha(x)$ are normalized  so that their
large $x$ expansions read
\bea
y_\alpha(x)= x^n(1-c_{\alpha,1}x^{-1}+c_{\alpha,2}x^{-2}
-c_{\alpha,3}x^{-3}+\cdots )\,.
\label{y_expansion}
\eea
The $1$-forms $d\log y_\alpha(x)$ are the direct analogs of Seiberg-Witten
differentials and define the chiral correlators by the conventional
contour integrals
\bea
\langle\tr \phi_\alpha^J\rangle =\oint_{\gamma_\alpha}\frac{dx}{2\pi i}
x^J\partial_x\log y_{\alpha}(x),
\label{trphiJ_contint}
\eea
where $\gamma_\alpha$ are large contours surrounding all zeros and poles of
$y_{\alpha}(x)$ in anti-clockwise
direction. Comparison of (\ref{y_expansion}), with (\ref{trphiJ_contint})
allows one to express the expansion coefficients $c_{\alpha,k}$ in terms
of chiral correlators $\langle\tr \phi_\alpha^J\rangle $ with $J\le k$. Here
are the first few relations
\bea
\label{trphiJ_vs_c}
&&\langle\tr \phi_\alpha\rangle=c_{\alpha,1}\nonumber\\
&&\langle\tr \phi_\alpha^2\rangle=c_{\alpha,1}^2-2c_{\alpha,2}\nonumber\\
&&\langle\tr \phi_\alpha^3\rangle=c_{\alpha,1}^3-3c_{\alpha,1}c_{\alpha,2}
+3 c_{\alpha,3}\\
&&\langle\tr \phi_\alpha^4\rangle=c_{\alpha,1}^4-4 c_{\alpha,1}^2 c_{\alpha,2}
+4 c_{\alpha,1}c_{\alpha,3}+2 c_{\alpha,2}^2-4 c_{\alpha,4}\nonumber\\
&&\qquad \cdots \qquad \cdots \qquad \cdots \qquad \cdots  \qquad
\cdots \qquad \cdots \nonumber
\eea
On the other hand, inserting the expansion (\ref{y_expansion}) into
(\ref{chi_alpha}) and comparing left and right hand sides one can express
the coefficients $c_{\alpha,k}$ (and due to (\ref{trphiJ_vs_c}) also
$\langle\tr \phi_\alpha^J\rangle$ ) in terms of coefficients of the
polynomials $\chi_\alpha(x)$. In fact the first $n$ of these relations can be
inverted to get the coefficients of the polynomials $\chi_\alpha(x)$ in
terms of $c_{\alpha,1},\ldots,c_{\alpha,n}$ (or, equivalently, in terms of
$\langle\tr \phi_\alpha\rangle, \ldots, \langle\tr \phi_\alpha^n\rangle$).
Then the remaining infinite number of relations are nothing but the
deformation of the celebrated chiral ring relations \cite{Cachazo:2002ry}
expressing higher power ($J>n$) chiral expectation values
$\langle\tr \phi_\alpha^J\rangle$ in terms of lower, up to the
$n$'th power expectation values\footnote{For the generalization
of chiral ring relation for the generic $\Omega$-background see
\cite{Fucito:2015ofa}.}.
For our later purposes let us display explicitly the relations
for the first three coefficients of the polynomials
\bea
\chi_\alpha(x)=\sum_{i=0}^n(-)^{i}\chi_{\alpha,i}x^{n-i}\,.
\eea
Expanding l.h.s. of (\ref{chi_alpha}) up to the order $\sim x^{n-3}$ we get
\bea
\label{chi_0}
&&\chi_{\alpha,0}=\sum_{1\le k_1<\cdots<k_{\alpha}\le r+1}
\prod_{\beta=1}^\alpha
\prod_{\gamma=\beta}^{k_\beta-1}q_\gamma\,\,;\\
\label{chi_1}
&&\chi_{\alpha,1}=\sum_{1\le k_1<\cdots<k_{\alpha}\le r+1}
\left(c_{0,1}-\sum _{\beta=1}^\alpha \left(c_{k_\beta-1,1}
-c_{k_\beta,1}\right)\right)
\prod_{\beta=1}^\alpha
\prod_{\gamma=\beta}^{k_\beta-1}q_\gamma\,\,;\\
\label{chi_2}
&&\chi_{\alpha,2}=\sum_{1\le k_1<\cdots <k_{\alpha}\le r+1}
\left[\sum _{\beta=1}^{\alpha-1} \sum _{\gamma=\beta+1}^\alpha
\left(c_{k_\beta-1,1}-c_{k_\beta,1}+
2\right) \left(c_{k_\gamma-1,1}-c_{k_\gamma,1}+2\right)\right.\nonumber\\
&&\qquad+\sum _{\beta=1}^\alpha \left(c_{k_\beta-1,1} \left(-c_{k_\beta,1}
+\alpha-\beta+3\right)
-\left(c_{0,1}+2 \alpha\right) \left(c_{k_\beta-1,1}-c_{k_\beta,1}+2\right)
\right. \nonumber\\
&&\left.\qquad\quad +(-\alpha+\beta-2) c_{k_\beta,1}
+c_{k_\beta-1,1}^2-c_{k_\beta-1,2}
+c_{k_\beta,2}+2 \alpha-2 \beta+3\right)\nonumber\\
&&\qquad\quad+\alpha \left(c_{0,1}
+\alpha\right)+c_{0,2}\bigg]
 \prod_{\beta=1}^\alpha
\prod_{\gamma=\beta}^{k_\beta-1}q_\gamma\,\,.
\eea
Clearly with more efforts it should be possible to write down expressions
for further coefficients but unfortunately these expressions soon become quite
intractable. In Section \ref{section4} we'll do
one more step giving an explicit expression for the next coefficient
in the special case when $r=1$.

Quite remarkably it is possible to eliminate the  functions
$y_2(x),\ldots ,y_r(x)$ from eq. (\ref{chi_alpha}) and find a single equation
for $y_1(x)$. Here is the result\footnote{Of course,
the same can be done also for $y_{r}(x)$.}:
\bea
1+\sum_{i=1}^{r+1}(-)^i\frac{\chi_i(x)}{y_1(x)}\prod_{j=1}^{i-1}
\frac{y_0(x-j)}{y_1(x-j)}\,q_j^{i-j}=0,
\label{y1eq}
\eea
It is useful to
represent the meromorphic functions $y_1(x)$ as a ratio:
\bea
y_1(x)=y_0(x)\,\,\frac{Y(x)}{Y(x-1)}\,,
\eea
where $Y(x)$ is an entire function with zeros located at
$x=a_{1,u}+(i-1)+\lambda_{1,u,i}$ (remind that $\lambda_{\alpha,u,i}$
is the appropriately rescaled length of the $i$'th row of
the Young diagram $Y_{\alpha,u}$).
In terms of
$Y(x)$ the eq. (\ref{y1eq}) can be rewritten as
\bea
\sum_{\alpha=0}^{r+1}(-)^\alpha\left(\prod_{\beta=1}^\alpha
q_\beta^{\alpha-\beta}\right)\chi_\alpha(x)Y(x-\alpha)=0\,.
\eea
Since for small values of the gauge
couplings $q_\alpha \ll 1$ the $i$'th row length \\ $\lambda_{\alpha,u,i}\rightarrow 0$ when  $i\rightarrow \infty$, it is reasonable to
expect that the sum
\bea
\psi(z)=\sum_{x\in {\mathcal Z}+a_{1,u}}Y(x)z^{-x}
\eea
will converge in some ring with the center located at $0$. Then the difference
equation for $Y(x)$ can be easily ``translated'' into a linear differential
equation
for $\psi(x)$ \cite{Fucito:2011pn,Fucito:2012xc}
\bea
\sum_{\alpha=0}^{r+1}(-)^\alpha\left(\prod_{\beta=1}^\alpha
q_\beta^{\alpha-\beta}\right)
\chi_\alpha(-z\frac{d}{dz})z^{-\alpha}\psi(z)=0
\label{diff_eq}
\eea
It is not difficult to find the coefficient in front of the highest derivative
$d^n/dz^n$ in (\ref{diff_eq}). Using (\ref{chi_0}) one can show that this
coefficient has a nice factorized form
\bea
\sum_{\alpha=0}^{r+1}(-)^{n+\alpha}\left(\prod_{\beta=1}^\alpha
q_\beta^{\alpha-\beta}\right)
\chi_{\alpha,0}z^{n-\alpha}
=(-)^nz^{n-r-1}\prod_{\alpha=0}^{r}\left(z-\prod_{\beta=1}^\alpha q_\beta\right)
\label{princ_coef}
\eea
Further investigation confirms that indeed (\ref{diff_eq}) is an $n$-th order
Fuchsian differential equation with $r+3$ regular singular points located at
\[
z_0=\infty,\,z_1= 1,\, z_2= q_1,\, z_3= q_1q_2,\,
\dots ,\,z_{r+1}= q_1q_2\cdots q_r,\,z_{r+2}=0,
\]
where for later use we have introduced new parameters $z_\alpha$ related
to the gauge couplings through conditions
\bea
q_\alpha=\frac{z_{\alpha+1}}{z_\alpha}\,.
\label{q_z_map}
\eea

\subsection{Exponents}
\label{exponents}
\subsubsection{Points $z_{r+2}=0$ and $z_0=\infty $}
First let's look after a solution of the form
\bea
\psi(z)=z^s(1+O(z))
\eea
Inserting this in (\ref{diff_eq}) we see that when $z\rightarrow 0$ the
term with $\alpha=r+1$ of (\ref{diff_eq}) is the most
singular one. So, for yet unknown constant $s$
we get the characteristic equation (sometimes called indicial equation)
\bea
y_{r+1}(r+1-s)=0
\label{ind_0}
\eea
Similarly the characteristic equation for the infinity reads
\bea
y_0(-s)=0
\label{ind_infty}
\eea

\subsubsection{Points $z_1=1$, $z_2=q_1$, $z_3=q_1q_2$, ... ,
$z_{r+1}=q_1q_2\cdots q_r$ }

Investigation of these points is slightly more subtle. Consider the ansatz
\bea
\psi(z)=\left(z-z_\alpha\right)^s\left(1
+O(z-z_\alpha))\right)
\eea
for some fixed $\alpha\in \{1,\ldots,r+1\}$. Taking into account
(\ref{princ_coef}) and (\ref{chi_1}) for the index $s$ we get the
equation
\bea
&&0=(-)^{n}z_\alpha^{n-r-1}\prod_{\beta\neq \alpha ,\beta=1}^{r+1}
(z_\alpha-z_\beta)\prod_{i=0}^{n-1}(s-i)\\
&&+\left(\sum_{\beta=0}^{r+1}n\beta(-z_\alpha)^{n-\beta-1}
\sum_{1\le k_1<\cdots<k_{\beta}\le r+1}\prod_{\gamma=1}^\beta z_{k_\gamma}\right)\prod_{i=0}^{n-2}(s-i)\nonumber\\
&&+\left(\sum_{\beta=0}^{r+1}(-z_\alpha)^{n-\beta-1}
\sum_{1\le k_1<\cdots<k_{\beta}\le r+1}\left(c_{0,1}-\sum_{\gamma=1}^\beta
\left(c_{k_{\gamma-1,1}}-c_{k_\gamma,1}\right)
\right)\prod_{\gamma=1}^\beta z_{k_\gamma}\right)\prod_{i=0}^{n-2}(s-i)\,. \nonumber
\eea
The first two lines come from the terms proportional to $(zd/dz)^n$.
The first (second) line includes part with $n$ "hits" ($n-1$ hits) on
$\psi(z)$ by the operator $d/dz$. The third line is coming from the terms
$\sim $ $(zd/dz)^{n-1}$ with all $n$ operators $d/dz$ hitting $\psi(z)$.
Though the second and especially third lines of this equation look quite
complicated, fortunately they can be simplified drastically.
Indeed it can be shown that the second line is equal to
\bea
(-)^n\,nz_\alpha^{n-r-1}
\prod_{\beta\neq \alpha ,\beta=1}^{r+1}
(z_\alpha-z_\beta)\prod_{i=0}^{n-2}(s-i)
\eea
while the third line can be rewritten as
\bea
(-)^nz_\alpha^{n-r-1}\left(c_{\alpha-1,1}-c_{\alpha,1}\right)
\prod_{\beta\neq \alpha ,\beta=1}^{r+1}
(z_\alpha-z_\beta)\prod_{i=0}^{n-2}(s-i)\,.
\eea
Thus for the allowed exponents we get
\bea
s\in\{0,1,\ldots n-2,c_{\alpha,1}-c_{\alpha-1,1}-1\}\,.
\eea
It is known in general that if indices differ by integers
one might have logarithmic solutions. This is not the case
however in the example at hand. A closer look ensures that
our differential equation around $z=z_\alpha$,
$\alpha=1,2,\ldots,r$ admits $n$ independent solutions of the types
\bea
&&(z-z_\alpha)^m+O((z-z_\alpha)^{n-1});
\quad m\in \{0,1,2,\ldots,n-2\}\quad
\text{and}\nonumber\\
&&(z-z_\alpha)^{c_{\alpha,1}-c_{\alpha-1,1}-1}(1+O(z-z_\alpha))\,.
\eea

%%%%%%%%%%%%%%%%%%%%%%%%%%%%%%%%%%%%%%%%%%%%%%%%%%%%%%%%%%%%%%%%%
%%%%%%%%%%%%%%%%%%%%%%%%%%%%%%%%%%%%%%%%%%%%%%%%%%%%%%%%%%%%%%%%%

\section{Toda CFT }
\label{section3}
\subsection{Preliminaries on $A_{n-1}$ Toda CFT}
These are 2d CFT theories which, besides the spin $2$ holomorphic
energy momentum current $W^{(2)}(z)\equiv T(z)$ are endowed with additional higher spin
$s=3\ldots,n $
currents $W^{(3)}$, \ldots $W^{(n)}$
\cite{Zamolodchikov:1985wn,Fateev:1987zh,Bilal:1988ze}. The Virasoro
central charge is conventionally parameterised as
\[
c=n-1+12 (Q,Q)\,,
\]
where the "background charge" $Q$ is given by
\[
Q=\rho (b+1/b)\,,
\]
where $\rho $ is the Weyl vector of the algebra $A_{n-1}$ and $b$ is the
dimensionless coupling constant of Toda theory. In what follows it would
be convenient to represent roots, weights and Cartan elements of $A_{n-1}$ as $n$-component vectors subject to condition that sum of
components is zero and endowed with the usual Kronecker scalar product.
Obviously this is equivalent
to a more conventional representation of these quantities as diagonal traceless
$n\times n$ matrices with pairing given by trace. In this representation
the Weyl vector is given by
\bea
\rho =\left(\frac{n-1}{2},\frac{n-3}{2},\ldots ,\frac{1-n}{2}\right)
\eea
and for the central charge we'll get
\[
c=(n-1)(1+n(n+1)q^2)
\]
where for the later use we have introduced the parameter
\[
q=b+\frac{1}{b}\,.
\]
For further reference let us quote here explicit expressions for the highest weight $\omega_1$ of the first fundamental representation and for its complete set of weights $h_1, \ldots,h_n$ ($h_1=\omega_1)$
\bea
&&(\omega_1)_k=\delta_{1,k}-1/n\,,\nonumber\\
&&(h_l)_k=\delta_{l,k}-1/n\,.
\eea
The primary fields $V_\alpha$ (in this paper we concentrate only on
the left moving holomorphic parts) are parameterized by vectors $\alpha $ with
vanishing center of mass. Their conformal wights are given by
\[
h_{\alpha}=\frac{(\alpha ,2Q-\alpha)}{2}\,.
\]
In what follows a special role is played by the fields $V_{\lambda \omega_1}$
with dimensions
\bea
h_{\lambda \omega_1}=\frac{\lambda(n-1)}{2} \left(q-\frac{\lambda}{n}\right)\,.
\label{dim_lambda}
\eea
For generic $\lambda$ these fields admit a single null vector at the first level.

%%%%%%%%%%%%%%%%%%%%%%%%%%%%%%%%%%%%%%%%%%%%%%%%%%%%%%%%%%%%%%%%%%%%%%%%%%%%
\subsection{Fusion with the completely degenerated field
$V_{-b\omega_1}$}
The field $V_{-b\omega_1}$ plays a special role in Toda theory.
Fusion rules with this field are especially simple (see e.g.
\cite{Fateev:2005gs})
\bea
V_{-b\omega_1}(z) V_{\alpha}(0)=\sum_{k=1}^nz^{b(Q_1-Q_k+\alpha_k)}
\left[V_{\alpha-bh_k}(0)\right],
\label{fusion_rule}
\eea
where $h_k$  are the weights of the first fundamental
representation and $[V_{\alpha}]$ denotes
the $W$-class of the primary field $V_{\alpha}$. In the case when
$V_{\alpha}$ is partially degenerated
(i.e. $ \alpha =\lambda \,\omega_1$ for some scalar $\lambda $ ),
then in (\ref{fusion_rule}) only the first two terms contribute,
all other classes drop out due to vanishing of the relevant
structure constants. The remaining exponents are equal to
respectively
\bea
\frac{\lambda b(n-1)}{n}  \qquad \text{and} \qquad  b\left(q-\frac{\lambda}{n}\right)\,.
\eea
In classical limit ($q\sim 1/b$)  the field $V_{-b\omega_1}$
satisfies the $n$-th order differential equation \cite{Fateev:2005gs,Fateev:2007ab}
\bea
\sum_{k=0}^nw^{(n-k)}(z)\partial_z^kV_{-b\omega_1}=0\,,
\label{diff_eq_W}
\eea
where $w^{(0)}\equiv 1$, $w^{(1)}\equiv 0$ and the other
coefficients $w^{(n-k)}(z)$ are the expectation values of
the currents $b^{n-k}W^{(n-k)}(z)$ in the classical limit
$b\rightarrow 0$ (the prefactor $b^{n-k}$ is included to
secure a finite limit).

%%%%%%%%%%%%%%%%%%%%%%%%%%%%%%%%%%%%%%%%%%%%%%%%%%%%%%%%%%%%%%%%%%%%%%%%%%%%%%%%%

\subsection{Derivation of null-vector decoupling equation in semiclassical
limit}

Let us consider the semi-classical limit of the correlator
\bea
\langle V_{\alpha^{(0)}}(\infty)V_{-b\omega_1}(z)
V_{\alpha^{(1)}}(z_1) \cdots V_{\alpha^{(r+1)}}(z_{r+1})
V_{\alpha^{(r+2)}}(0)\rangle\,,
\label{correlator}
\eea
where we'll assume that all the fields besides $V_{-b\omega_1}(z)$
are "heavy", namely
\bea
\alpha^{(0)}=\eta^{(0)}/b; \qquad\qquad
\alpha^{(r+2)}=\eta^{(r+2)}/b; \qquad\qquad
\alpha^{(k)}=\eta^{(k)}\omega_1/b
\eea
(the parameters $\eta^{(0)}$ and $\eta^{(r+2)}$ are $n$-component
vectors while $\eta^{(k)}$ are scalars all of them remaining finite in
the $b\rightarrow 0$ limit). We have chosen all the parameters, besides
the first and the last ones, to be proportional to the first fundamental
weight $\omega_1$ since this is the case when the AGT relation between
correlation functions and the partition function of quiver gauge
theory holds. As we'll see later, the AGT correspondence in the $b\rightarrow 0$ limit emerges as a special case of a more general construction presented in the remaining part of this
paper.  In semiclassical limit the correlator (\ref{correlator})
factorizes into a product of the classical (normalized) expectation
value of the light field $V_{-b\omega_1}(z)$ with the
correlator of the remaining heavy operators, hence, with respect
to the variable $z$ it must satisfy the same differential equation (\ref{diff_eq_W}):
\bea
\left(\frac{d^n}{dz^n}+\sum_{k=2}^nw^{(k)}(z)\frac{d^{n-k}}
{dz^{n-k}}\right)G(z)=0\,.
\label{diff_eq_W_final}
\eea
As already mentioned the coefficient functions
$w^{(n-k)}(z)$ are the classical expectation values of the holomorphic
currents $b^{n-k}W^{(n-k)}(z)$ in the background of heavy
operators. Due to the form of OPE of
$W^{(k)}$-current with primary fields, this classical expectation
values should be rational functions of $z$, with $k$-th order poles
located at the insertion points of heavy primary fields. The OPE
(\ref{fusion_rule}) completely fixes the indices at the singular
points as follows:
\bea
&&z=0; \,\,\,\quad u-1+(\eta^{(r+2)})_u \quad \text{for}\quad
u=1,2,\ldots,n \nonumber\\
&&z=z_{\alpha};\quad 1-\eta^{(\alpha)}/n\quad \text{and} \quad
(n-1)\eta^{(\alpha)}/n\quad \text{for}\quad
\alpha=1,2,\ldots,r+1\quad\nonumber\\
&&z=\infty; \quad u-1+(\eta^{(0)})_u \quad \text{for}\quad u=1,2,\ldots,n\,.
\label{w_deq_indices}
\eea
There is a small puzzle to be understood here. The indicial equation at
$z=z_\alpha$ is a degree $n$ algebraic equation while on the second
line of eq. (\ref{w_deq_indices}) we quoted only two indices. Multiple
roots are not admissible, both field appearing on the r.h.s. of  the OPE
should come with multiplicity one. To find the missing indices, let us
slightly change the charge parameters of the field at the point $z_\alpha$. We'll
immediately see that besides two indices close to those given on the second line of
(\ref{w_deq_indices}), there are $n-2$ additional indices, located close to the points
\bea
2-\eta^{(\alpha)}/n,\,3-\eta^{(\alpha)}/n,\,\ldots\, ,n-1-\eta^{(\alpha)}/n\,.
\eea
Thus it is natural to assume that besides $(n-1)\eta^{(\alpha)}/n$ we have
sequence of $n-1$ indices
\bea
1-\eta^{(\alpha)}/n,\,2-\eta^{(\alpha)}/n,\,\ldots\, ,n-1-\eta^{(\alpha)}/n\,.
\eea
The extra indices we got are naturally attributed to the contribution of
descendant fields.
This is not the end of story yet. It is well known
that if there are indices at a singular point which differ from
each other by integers, generally speaking logarithmic solutions
emerge, something, which is not acceptable in a standard CFT such
as Toda theory. The condition that logarithmic solutions are actually
absent, imposes further restrictions on the rational coefficient
functions $w^{(k)}(z)$. We'll explicitly parameterize these rational
functions as
\bea
w^{(k)}(z)=\sum_{\alpha=1}^{r+2}\sum_{m=0}^{k-1}{\frac{w^{(k,\alpha)}_{-m}}
{(z-z_\alpha)^{k-m}}}\,.
\label{wk_function}
\eea
Consider a solution of the differential equation (\ref{diff_eq_W_final}) around $z=z_\alpha$
represented as a series
\bea
G(z)=\sum_{l=0}^\infty b_l (z-z_\alpha)^{l+s_\alpha}\,.
\eea
Let us insert this expansion into (\ref{diff_eq_W}) and read off
the first $n-1$ constraints imposed by the differential equation. We immediately
get the relations
\bea
b_m (s_\alpha+m-n+1)_n+\sum_{k=2}^n\sum_{l=m-k+1}^m b_lw^{(k,\alpha)}_{m-l}
(s_\alpha+l-n+k+1)_{n-k}=0
\label{ind_eq_gen}
\eea
valid for $m\in \{0,1,\ldots, n-2\}$, where
\bea
(x)_l=x(x+1)\cdots (x+l-1)
\eea
is the Pochhammer's symbol. According to what has been discussed above,
the coefficients $w^{(n-k)}(z)$ of the differential equation should be chosen so that
these constraints be satisfied for
\bea
s_\alpha=1-\frac{\eta^{(\alpha)}}{n}
\label{s_eta}
\eea
and for arbitrary constants $b_0, \ldots,b_{n-2}$. In particular choosing
$b_l=\delta_{l,m}$ we arrive at
\bea
(s_\alpha+m-n+1)_n+\sum_{k=2}^nw^{(k,\alpha)}_{0}
(s_\alpha+m-n+k+1)_{n-k}=0\,,
\eea
satisfied for each $m=0,1,\ldots ,n-2$.
Solving this system of equations with respect to $w^{(k,\alpha)}_{0}$
we get
\bea
w^{(k,\alpha)}_{0}=(-1)^{k+1}
\left(
\begin{array}{c}
n\\
k
\end{array}
\right)
(k-1)(s_\alpha-1)_k\,,
\label{w0_vs_s}
\eea
where the standard notation
\bea
\left(
\begin{array}{c}
n\\
k
\end{array}
\right)=\frac{n!}{k!(n-k)!}\nonumber
\eea
for the binomial coefficient is used.
It is rewarding to see that this formula with parameter $s_\alpha$ specified
in (\ref{s_eta}) gives correct zero mode eigenvalues of W-currents
on the field $V_{\omega_1 \eta_{\alpha}/b}$ in semiclassical limit
$b\rightarrow 0$. In particular for the (rescaled) conformal dimension
(i.e. for $k=2 $) we get
\bea
w^{(2,\alpha)}_{0}=\frac{(n-1)\eta^{(\alpha)}
\left(1-\eta^{(\alpha)}/n\right)}{2}
\label{dim_eta}
\eea
in complete agreement with (\ref{dim_lambda}). I have checked that also the
other eigenvalues agree with data available in the
literature.\\
The remaining constraints that follow from eq. (\ref{ind_eq_gen})
can be represented as
\bea
\sum_{k=m+1}^nw^{(k,\alpha)}_{-m}(s_\alpha-n+k+l)_{n-k}=0, \qquad\text{for}
\quad l=1,2,\ldots,n-m-1
\eea
valid for each value of $m\in \{1,2,\ldots,n-2\}$.
These equations allow one to express all quantities
$w^{(k,\alpha)}_{-m}$ with $k\in\{2,3,\ldots, n\}$
and $m\in \{1,2,\ldots,k-1\}$
in terms of $w^{(m+1,\alpha)}_{-m}$. Here is
the explicit expression:
\bea
w^{(k,\alpha)}_{-m}=(-1)^{k-m+1}
\left(
\begin{array}{c}
n-m-1\\
n-k
\end{array}
\right)
(s_\alpha)_{k-m-1}w^{(m+1,\alpha)}_{-m}\,.
\label{W_vs_basic_W}
\eea
Thus we managed to express all the coefficients of higher order
poles at $z=z_\alpha $, $\alpha=1,2,\ldots,r+1$
(but not those corresponding to $z=0$) in terms of
the residues $w^{(m+1,\alpha)}_{-m}$.

Recall now that there should be one more independent index at
$z=z_{\alpha}$
(see middle line of (\ref{w_deq_indices}) and (\ref{s_eta}))
equal to
\bea
\tilde{s}_\alpha=(n-1)\eta^{(\alpha)}/n=(n-1)(1-s_\alpha) \,.
\label{s_tilde}
\eea
The indicial equation to be satisfied is (cf. (\ref{ind_eq_gen})):
\bea
(\tilde{s}_\alpha-n+1)_n+\sum_{k=2}^nw^{(k,\alpha)}_{0}
(\tilde{s}_\alpha-n+k+1)_{n-k}=0\,.
\label{ind_s_tilde}
\eea
It can be checked that quite remarkably this equality
under substitutions (\ref{w0_vs_s}), (\ref{s_tilde})
is indeed satisfied automatically.

It remains to take into account constraints imposed by Ward identities.
These identities are consequences of \footnote{The sign factor
$(-1)^{k+1}$ reflects the fact that a two-point function of
primaries is non-zero if the zero modes of even spin currents
(e.g. dimensions) coincide while those of odd currents have
opposite signs (see e.g. \cite{Fateev:2007ab}).}
\bea
\langle \alpha^{(0)}|W^{(k)}_m=(-1)^{k+1}w^{(k,0)}_0\delta_{m,0};
\qquad m\ge 0
\label{bra_state_W}
\eea
and the commutation relations
\bea
[W^{(k)}_m,V_{\alpha}(z)]=\sum_{l=1}^kz^{k+m-l}
\left(
\begin{array}{c}
k+m-1\\
l-1
\end{array}
\right)
W^{(k,z)}_{l-k} V_{\alpha}(z)\,,
\label{W_V_commutator}
\eea
where
\bea
W^{(k,z)}_l V_{\alpha}(z)=\oint_z(\zeta-z)^{l+k-1}W^{(k)}(\zeta)
V_{\alpha}(z)\frac{d\zeta}{2\pi i}
\label{desc_field}
\eea
is a descendant field on the level $-l$ for $l<0$. If $l>0$,
(\ref{desc_field}) vanishes, while for $l=0$
\bea
W^{(k,z)}_0 V_{\alpha}(z)=w^{(k)}_0 V_{\alpha}(z)\,,
\eea
$w^{(k)}_0$ being the zero mode eigenvalue corresponding to
the field $V_{\alpha}$.
For $m>0$ combining (\ref{bra_state_W}) and (\ref{W_V_commutator}) we
easily get
\bea
w^{(k,r+2)}_{-m}=-\sum_{\alpha=1}^{r+1}\sum_{l=1}^{k-m}
z_\alpha^{k-m-l}\binom{k-m-1}{l-1}w^{(k,\alpha)}_{l-k}
\eea
or, in view of (\ref{W_vs_basic_W})
\bea
w^{(k,r+2)}_{-m}=\sum_{\alpha=1}^{r+1}\sum_{l=1}^{k-m}
(-)^{l}z_\alpha^{k-m-l}\binom{k-m-1}{l-1}\binom{n+l-k-1}{l-1}
(s_\alpha)_{l-1}w^{(k-l+1,\alpha)}_{l-k}\,.\qquad
\label{W_last_vs_basic}
\eea
Similarly analysing $m=0$ case of (\ref{bra_state_W})
and (\ref{W_V_commutator}) we get the relation
\bea
&&(-1)^{k+1}w^{(k,0)}_{0}+\sum_{\alpha=1}^{r+2}
w^{(k,\alpha)}_{0}\nonumber\\
&&+\sum_{\alpha=1}^{r+1}
\sum_{l=1}^{k-1}(-)^{l+1}z_\alpha^{k-l}\binom{k-1}{l-1}
\binom{n+l-k-1}{l-1}
(s_\alpha)_{l-1}w^{(k-l+1,\alpha)}_{l-k}=0\,.
\label{w1_walpha_system}
\eea
These equations allow one to express all coefficients
related to the point $z_1=1$ (i.e. $w^{(k,1)}_{1-k}$)
in terms of coefficients $w^{(k,\alpha)}_{1-k}$ with
$\alpha=2,3,\ldots,r+1$. Indeed the system (\ref{w1_walpha_system})
with respect to the variables $w^{(k,1)}_{1-k}$ has a Gaussian
triangular form and can be solved. As a side remark note that
this problem reduces to the inversion of the $(n-1)\times (n-1)$ lower
triangular matrix
\bea
M_{i,j}=\frac{(s)_{i-j}}{((i-j)!)^2}\,,
\eea
where $i,j\in \{1,2,\ldots,n-1\}$. It's possible to show that
the inverse matrix can be represented as
\bea
(M^{-1})_{i,j}=\sum_{l=0}^{n-2}P_l(s)\delta_{i-j,l}\,,
\eea
where the polynomials $P_l(s)$ are conveniently given by means of a generating function
\bea
\sum_{l=0}^{\infty} P_l(s)q^l =
\frac{1}{_1F_1(s,1,q)}
\eea
with
\bea
_1F_1(a,b,q)=\sum_{l=0}^\infty\frac{(a)_l}{(b)_ll!}q^l\,.
\eea

For the later reference let us write down $w^{(2)}(z)$
(see eq. (\ref{wk_function})) in terms of dimensions and
the parameters $w_{-1}^{(2,\alpha)}$,
$\alpha=2,3,\ldots,r+1$, explicitly
\bea
w^{(2)}(z)&=&\frac{w_{0}^{(2,r+2)}}{z^2}+
\sum_{\alpha=1}^{r+1}\frac{w_{0}^{(2,\alpha)}}{(z-z_\alpha)^2}
\nonumber\\
&+&\frac{w_{0}^{(2,0)}-\sum_{\alpha=1}^{r+2}
w_{0}^{(2,\alpha)}}{z(z-z_1)}
+\sum_{\alpha=2}^{r+1}\frac{z_\alpha(z_\alpha-z_1)
w_{-1}^{(2,\alpha)}}{z(z-z_1)(z-z_\alpha)}\,.
\label{w2_function}
\eea
To conclude we succeeded to express all the parameters of the
differential equation (\ref{diff_eq_W_final}) in terms of (see eqs. (\ref{w0_vs_s}),
(\ref{W_vs_basic_W}), (\ref{W_last_vs_basic}) and
(\ref{w1_walpha_system})):
\begin{itemize}
\item{zero-mode eigenvalues of the $W$-currents corresponding to
the initial state, insertion fields and the final state
(i.e. $w^{(k,\alpha)}_{0}$ for $k\in\{2,3,\ldots,n\}$,
$\alpha\in \{0,1,\ldots, r+2\}$)}
\item{the coefficients $w^{(k,\alpha)}_{1-k}$
for$k\in\{2,3,\ldots,n\}$, $\alpha\in\{2,3,\ldots,r+1\}$,
which are residues of rational functions
$w^{(k)}(z)$ at the points $z=z_\alpha$}
\end{itemize}
The parameters of the second list will be referred as the accessory
parameters, since these are direct generalizations of the accessory
parameters of the Liouville theory (the particular $n=2$ case of Toda theory) \cite{Takhtajan:1985}. To avoid confusion notice that we have in mind not
the "real" monodromy problem of \cite{Takhtajan:1985},
but a generalization of the complex $SL(2,\mathbb{C})$ monodromy
problem, discussed in \cite{Nekrasov:2011bc,Litvinov:2013sxa}, to the
$SL(n,\mathbb{C})$ case.

It is essential to note that there are exactly $(n-1)\times r$
accessory parameters, as many as the number of parameters necessary
to fix $r$ intermediate $W$-families of the conformal block.
We will see soon, that the DSW "curve" is the appropriate tool
to solve the related monodromy problem, namely, to find such
set of accessory parameters, which corresponds to a given
set of intermediate $W$-families. In particular concentrating on
the accessory parameters $w^{(2,\alpha)}_{-1}$, we will easily
reestablish the famous AGT relations (in the semiclassical limit).
Consideration of the remaining accessory parameters lead to a generalization
of AGT, relating conformal blocks including non-primary fields on
2d CFT side with gauge invariant expectation values on the ${\cal N}=2$
gauge theory side. Investigation of this new relations in general
quantum Toda case (e.g. along the lines of
\cite{Bourgine:2015szm,Nekrasov:2015wsu}, seems to be quite an interesting task.

%%%%%%%%%%%%%%%%%%%%%%%%%%%%%%%%%%%%%%%%%%%%%%%%%%%%%%%%%%%%%%%%%%%%%%%%%%%%
%%%%%%%%%%%%%%%%%%%%%%%%%%%%%%%%%%%%%%%%%%%%%%%%%%%%%%%%%%%%%%%%%%%%%%%%%%%%

\section{Comparison with the differential equation
derived from DSW}
\label{section4}
%%%%%%%%%%%%%%%%%%%%%%%%%%%%%%%%%%%%%%%%%%%%%%%%%%%%%%%%%%%%%%%%%%%%%%%%%%%%
\subsection{From the gauge theory differential equation to
the null-vector decoupling equation}
We'll argue here that the differential equation obtained
from (\ref{diff_eq}) by substitution
\bea
\psi(z)= G(z)\prod_{\alpha=1}^{r+2}
(z-z_\alpha)^{t_\alpha}
\label{gauge_subst}
\eea
with suitably chosen parameters $t_\alpha$ coincides with the
differential equation (\ref{diff_eq_W_final}) satisfied by Toda CFT semi-classical
conformal block (\ref{correlator}). The idea is to choose
parameters $t_\alpha$ so
that the term with derivative of order $n-1$ disappears.
Straightforward calculations show that this task is achievable
with the unique choice
\bea
t_{r+2}&=&r+1+\frac{1-n}{2}-\frac{c_{r+1,1}}{n}\,;\nonumber\\
t_{\alpha}&=&\frac{c_{\alpha,1}-c_{\alpha-1,1}}{n}-1, \qquad
\alpha=1,2,3,\ldots ,r+1\,.
\eea
The resulting differential equation for $G(z)$ can be represented
as
\bea
\left(\frac{d^n}{dz^n}+S_2(z)\frac{d^{n-2}}{dz^{n-2}}+
\cdots +S_n(z)\right)G(z)=0\,,
\label{diff_eq_G}
\eea
where, in particular,
\bea
S_2(z)&=&\left(\frac{n\left(n^2-1\right)}{24}
-\frac{n-1}{2 n}\,\, c_{r+1,1}^2+c_{r+1,2}\right)
\frac{1}{z^2}\nonumber\\
&+&\sum_{\alpha=1}^{r+1}\left(\frac{(n-1) \left(c_{\alpha ,1}
-c_{\alpha -1,1}\right) \left(c_{\alpha -1,1}
-c_{\alpha ,1}+n\right)}{2n(z-z_\alpha)^2}\right.\nonumber\\
&+&\left. \frac{c_{\alpha -1,2}-c_{\alpha ,2}
+\left(c_{\alpha -1,1}-c_{\alpha ,1}\right) \left(-c_{\alpha ,1}+\frac{c_{r+1,1}}{n}+\alpha +\frac{n-3}{2}
-r\right)}{z(z-z_\alpha)}\right)\nonumber\\
&+&\sum _{1\le\alpha<\beta \le r+1}
 \frac{\left(c_{\alpha -1,1}-c_{\alpha ,1}\right)
  \left(c_{\beta -1,1}-c_{\beta ,1}+n\right)}
  {n \left(z-z_{\alpha }\right) \left(z-z_{\beta }\right)}\,.
\label{S2_function}
\eea
In general $S_u(z)$ for $u=2,3,\ldots,n$, are rational functions
with poles of order $u$ located at
$z\in\{z_1=1,z_2,\ldots,z_{r+1},z_{r+2}=0\}$.

Note that the substitution (\ref{gauge_subst}) shifts all exponents
described in Sec. \ref{exponents} in an obvious manner. Namely
the exponents at $z_\alpha$ get shifted by $t_\alpha$ for
$\alpha=1,2,\ldots,r+2$ and the exponent at $z_0=\infty$
by
\[
t_{r+2}+\cdots+t_{r+1}=\frac{1-n}{2}-\frac{c_{0,1}}{n}\,,
\]
so that the resulting indices become
\bea
&&z=0; \quad \frac{n-1}{2}-a_{r+1,u}+\frac{c_{r+1,1}}{n},
\quad \quad u=1,2,\ldots,n \nonumber\\
&&z=z_{\alpha};\quad u-\frac{c_{\alpha,1}-c_{\alpha-1,1}}{n}
\quad \text{and} \quad
\frac{(n-1)(c_{\alpha,1}-c_{\alpha-1,1})}{n},\quad
u=1,2,\ldots,n-1\qquad\nonumber\\
&&z=\infty; \quad \frac{n-1}{2}-a_{0,u}+\frac{c_{r+1,1}}{n},
\quad u=1,2,\ldots,n\,.
\label{DSW_deq_indices}
\eea
Remind that $a_{0,u}$ and $a_{r+1,u}$
are the roots of the polynomials $y_0(x)$ and $y_{r+1}(x)$
respectively (see eqs. (\ref{y0_yrp1_factorized})). In particular
\bea
\sum_{u=1}^na_{0,u}=c_{0,1}\quad \text{and} \quad
\sum_{u=1}^na_{r+1,u}=c_{r+1,1}\,.
\eea
Now we see that the simple identification
\bea
\frac{n+1}{2}-u-(\eta^{(r+2)})_u&\equiv& a_{r+1,u}
-\frac{c_{r+1,1}}{n}\,;\nonumber\\
\eta^{(\alpha)}&\equiv& c_{\alpha,1}-c_{\alpha-1,1}\,;\nonumber\\
\frac{n+1}{2}-u-(\eta^{(0)})_u&\equiv& a_{0,u}-\frac{c_{0,1}}{n}
\label{AGT_map}
\eea
provides a perfect matching of indices (\ref{DSW_deq_indices})
with (\ref{w_deq_indices}).

%%%%%%%%%%%%%%%%%%%%%%%%%%%%%%%%%%%%%%%%%%%%%%%%%%%%%%%%%%%%%%%%%%%%
\subsection{Matching $S_2(z)$ with $w^{(2)}(z)$: emergence of AGT}
Now let us compare $S_2(z)$ (eq. (\ref{S2_function})) with
$w^{(2)}(z)$ (eq. (\ref{w2_function})). Clearly the
coefficients at the double poles under the map (\ref{AGT_map})
become  identical (not a surprise, since we have already
checked that indices coincide).

Comparison of residues of poles at $z=z_\alpha$ for $\alpha\in
\{2,\ldots,r+1\}$ leads to identification
\bea
&&w_{-1}^{(2,\alpha)}= \frac{c_{\alpha -1,2}-c_{\alpha ,2}
+\left(c_{\alpha -1,1}-c_{\alpha ,1}\right) \left(-c_{\alpha ,1}
+\frac{c_{r+1,1}}{n}+\alpha +\frac{n-3}{2}
-r\right)}{z_\alpha}\nonumber\\
&&+\sum_{\beta=1}^{\alpha-1}\frac{\left(c_{\beta -1,1}
-c_{\beta ,1}\right)\left(c_{\alpha -1,1}-c_{\alpha ,1}+n\right)}
{n\left(z_\alpha-z_\beta\right)}
+\sum_{\beta=\alpha+1}^{r+1}\frac{\left(c_{\alpha -1,1}
-c_{\alpha ,1}\right)\left(c_{\beta -1,1}-c_{\beta ,1}+n\right)}
{n \left(z_\alpha-z_{\beta }\right)}\,.\nonumber\\
\label{w2_res_z_alpha}
\eea
Matching the residue at the pole $z=z_1\equiv 1$ requires
\bea
&&w_{0}^{(2,0)}-\sum_{\alpha=1}^{r+2}w_{0}^{(2,\alpha)}
-\sum_{\alpha=2}^{r+1}z_\alpha w_{-1}^{(2,\alpha)}
=\nonumber\\
&&c_{0,2}-c_{1,2}+\left(c_{0,1}-c_{1,1}\right)
\left(-c_{1,1}+\frac{c_{r+1,1}}{n}+\frac{n-1}{2}-r\right)\nonumber\\
&&\qquad\qquad\qquad\qquad+\sum_{\alpha=2}^{r+1}\frac{\left(c_{0,1}-c_{1,1}\right)
\left(c_{\alpha -1,1}-c_{\alpha ,1}+n\right)}
{n\left(1-z_\alpha\right)}\,.
\label{w2_res_z_1}
\eea
A careful calculation shows that upon replacement of dimensions
$w_{0}^{(2,\alpha)}$ by their expressions in terms of parameters
$\eta^{(\alpha)}$ specified in (\ref{AGT_map}) and substitution
(\ref{w2_res_z_alpha}) for $w_{-1}^{(2,\alpha)}$, the eq.
(\ref{w2_res_z_1}) becomes an identity. To prove this statement
we have used the useful identity (for our purpose one should take
$d_\gamma =c_{\gamma,1}-c_{\gamma-1,1}$)
\bea
&&\sum_{\alpha=2}^{r+1}z_\alpha\left(
\sum_{\beta=1}^{\alpha-1}\frac{d_\beta\left(n-d_\alpha\right)}
{n\left(z_\alpha-z_\beta\right)}
+\sum_{\beta=\alpha+1}^{r+1}\frac{d_\alpha\left(n-d_\beta\right)}
{n \left(z_\alpha-z_{\beta }\right)}\right)\nonumber\\
&&=\int_{\cal{C}}\sum_{1\le\beta<\gamma\le r+1}
\frac{z d_\beta(n-d_\gamma)}{(z-z_\beta)(z-z_\gamma)}
\,\,\frac{dz}{2\pi i}\nonumber\\
&&=\sum_{1\le\beta<\gamma\le r+1}
d_\beta(n-d_\gamma)-\sum_{\gamma=2}^{r+1}
\frac{d_1(n-d_\gamma)}{z_1-z_\gamma}
\eea
where the contour $\cal{C}$ encloses the points $z_2,\ldots,z_{r+1}$
but not $z_1$. To pass from the second line to the third, one
should notice that the same integral alternatively can be
computed as sum of residues at infinity and at $z_1$ taken
with negative signs.

There is no need to compare residues at the remaining pole at
$z=z_{r+2}=0$ since all residues sum to zero (both
$w^{(2)}(z)$ and $S_2(z)$ vanish at large values of $z$
as $1/z^2$).

Now let us look on identification (\ref{w2_res_z_alpha})
more closely.
The gauge theory expectation value
\bea
\langle\tr \phi_\alpha^2\rangle=c_{\alpha,1}^2-2c_{\alpha,2}
\label{trphi2_vs_c}
\eea
is related to the partition function through Matone relation
\cite{Matone:1995rx} which is valid also in presence of a nontrivial
$\Omega$-background \cite{Flume:2004rp}:
\bea
\langle\tr \phi_\alpha^2\rangle=\sum_{u=1}^na_{\alpha,u}^2
-b^{2}q_\alpha\partial_{q_\alpha}\log Z_{inst}\,
\label{Matone}
\eea
where $Z_{inst}$ is the instanton part of the partition
function and $a_{\alpha,u}$ are the Coulomb branch parameters
which are related to the parameters $c_{\alpha,1}$ by
\bea
\sum_{u=1}^na_{\alpha,u}=c_{\alpha,1}
=\langle\tr \phi_\alpha\rangle\,.
\label{c1_vs_a}
\eea
Notice also that in (\ref{Matone}) taking into account our
  convention $\epsilon_2=1$ we have replaced
\bea
\epsilon_1\epsilon_2=
\frac{\epsilon_1}{\epsilon_2}=b^2\,,
\eea
where the last equality, as we'll see soon, is necessary
to match the gauge theory and Toda theory sides.
It will be convenient to separate the "center of mass" of
the quantities $a_{\alpha,u}$ introducing new parameters (these are
parameters indicated in Fig.\ref{quiv_block}\cal{b})
\bea
P_{\alpha,u}=a_{\alpha,u}-\frac{c_{\alpha,1}}{n}
\label{a_cm_dropped}
\eea
Let me emphasize that the quantities $a_{\alpha,u}$
are genuine parameters of our gauge theory and
as such, they do not depend on the gauge couplings
$q_\alpha$ (or, equivalently, on $z_\alpha$).
Thus, on r.h.s. of eq. (\ref{w2_res_z_alpha}), besides their
explicit appearance, the variables $z_\alpha$ are hidden
only in the differences $c_{\alpha -1,2}-c_{\alpha ,2}$. Due to
(\ref{q_z_map}) we have
\bea
q_\alpha\partial_{q_\alpha}=\sum_{\beta=\alpha+1}^{r+1}
z_\beta\partial_{z_\beta}\,.
\eea
Hence, combining (\ref{trphi2_vs_c}), (\ref{Matone}),
(\ref{c1_vs_a}) and (\ref{a_cm_dropped}) we get
\bea
c_{\alpha -1,2}-c_{\alpha ,2}=\frac{1}{2}\sum_{u=1}^n
\left(P_{\alpha,u}^2-P_{\alpha-1,u}^2\right)
&+&\frac{n-1}{2n}\left(c_{\alpha -1,1}^2
-c_{\alpha ,1}^2\right)\nonumber\\
&+&b^2 z_\alpha\partial_{z_\alpha}\log Z_{inst}\,.
\eea
Inserting this expression into (\ref{w2_res_z_alpha})
after few simple manipulations, for \\$\alpha\in\{2,\ldots,r+1\}$
we obtain
\bea
&&w_{-1}^{(2,\alpha)}=b^2 \partial_{z_\alpha}\log Z_{inst}
+\left(h(P_{\alpha-1,u})-h(P_{\alpha,u})
-w_{0}^{(2,\alpha)}\right)
\partial_{z_\alpha}\log z_\alpha\nonumber\\
&&-\sum_{\beta=1}^{\alpha-1}\eta^{(\beta)}
\left(1-\frac{\eta^{(\alpha)}}{n}\right)\partial_{z_\alpha}
\log \left(1-\frac{z_\alpha}{z_\beta}\right)
-\sum_{\beta=\alpha+1}^{r+1}\eta^{(\alpha)}
\left(1-\frac{\eta^{(\beta)}}{n}\right)
\partial_{z_\alpha}\log \left(1-\frac{z_\beta}
{z_\alpha}\right)\,, \nonumber\\
\label{wm1_final}
\eea
where
\bea
h(P_{\alpha,u})=\frac{n(n^2-1)}{24}-\frac{1}{2}
\sum_{u=1}^nP_{\alpha,u}^2
\eea
are the (rescaled) dimensions with Toda momenta $P_{\alpha,u}$
(Toda momenta are related to the charge parameters via
$\eta=\rho-P$) and dimensions $w_{0}^{(2,\alpha)}$ are given
in (\ref{dim_eta}). Remind now that $w_{-1}^{(2,\alpha)}$ is
related to the Virasoro generator $b^{-2}L_{-1}$ acting on the
field $V_{\lambda_\alpha}(z_\alpha)$. Thus
\bea
w_{-1}^{(2,\alpha)}=b^2\partial_{z_\alpha}\log F_{CFT}\,,
\label{wm1_vs_block}
\eea
where $F_{CFT}$ is the $r+3$ point conformal block of Toda theory
depicted in Fig.\ref{quiv_block}{\cal b}.
Comparing (\ref{wm1_final}) with (\ref{wm1_vs_block}) it is not hard
to recognize the celebrated AGT relation.

The same technique, in principle, can be applied
to investigate the remaining accessory parameters
$w_{1-k}^{(k,\alpha)}$ with $k\in \{3,\ldots,n\}$.
The result would be an extension of the AGT correspondence to the
case of conformal blocks including a descendant field on CFT side and
higher power expectation values in gauge theory side.
Unfortunately in general case the next steps along this line
seem to be quite complicated, one soon encounters awkward expressions.
Instead in the next section we will give few more details for the
simpler case of four point functions (i.e. for $r=1$).

%%%%%%%%%%%%%%%%%%%%%%%%%%%%%%%%%%%%%%%%%%%%%%%%%%%%%%%%%%%%%%%%%%%%%%%%%%%%

\subsection{$A_{n-1}$-Toda $4$-point functions versus $SU(n)$
gauge theory with $2n$ fundamental hypers}
In this case we have a single DSW curve equation
\bea
\chi_1(x)=\frac{q_1 y_0(x-1) y_2(x)}{y_1(x-1)}+y_1(x)
\eea
The first three coefficients of the $n$'th order polynomial
$\chi_1(x)$ can be read off from the general formulae
(\ref{chi_0})-(\ref{chi_2}). We'll need also the forth
coefficient which in our case $r=1$ is easy to obtain. Without
loss of generality we can set\footnote{In fact a uniform
shift of all parameters $a_{\alpha,u}$ indicated in
Fig.\ref{quiv_block}\cal{a} is immaterial and can be compensated by
the shift of same amount of the parameter $x$ of DSW equations (\ref{chi_alpha}).} $c_{1,1}=0$. As a result we get
\bea
\chi_{1,0}&=&1+q_1; \qquad \chi_{1,1}=q_1 \left(c_{0,1}+c_{2,1}\right);
\nonumber\\
\chi_{1,2}&=&c_{1,2}+q_1 \left(c_{0,2}-c_{1,2}+c_{0,1} \left(c_{2,1}-1\right)+c_{2,2}\right);\nonumber\\
\chi_{1,3}&=&c_{1,3}-q_1 \left(c_{0,3}
+2 c_{1,2}+c_{1,3}+c_{0,2} \left(c_{2,1}-2\right)-c_{1,2} c_{2,1}\right.\nonumber\\
&&\hspace{2.2cm}\left. +c_{0,1} \left(-c_{1,2}-c_{2,1}+c_{2,2}+1\right)
+c_{2,3}\right)\,.
\label{Chi_0123_special}
\eea
Now starting from the eq. (\ref{diff_eq}) and following
the steps described in previous sections  we'll try to recast the
differential equation in the form (\ref{diff_eq_G}). The
coefficient function $S_2(z)$ of (\ref{diff_eq_G}) we
have already calculated for the general case, so we'll
simply specify the expression (\ref{S2_function}) to the
case $r=1$.

The data (\ref{Chi_0123_special}) are sufficient for
calculation of the next coefficient function $S_3(z)$ of
(\ref{diff_eq_G}). The calculation is rather lengthy but
quite straightforward. Here we'll not give all the details.
Those interested readers who might have desire to recover
the results presented below  could benefit from the useful
identity
\bea
\left(z\frac{d}{dz}\right)^n=\sum_{l=1}^nd_{n,l}z^l
\left(\frac{d}{dz}\right)^l\,,
\eea
where the expansion coefficients admit a nice
representation as
\bea
d_{l+k,l}=\sum_{i_1=1}^{l}\sum_{i_2=1}^{i_1}\cdots
\sum_{i_l=1}^{i_{l-1}}i_1i_2\cdots i_{l}\,.
\label{diff_exp_gen}
\eea
In fact, to calculate $S_3(z)$ one needs only the
first four coefficients which can be easily deduced
from (\ref{diff_exp_gen})
\bea
&&d_{n,1}=1\,;\nonumber\\
&&d_{n,2}=\frac{n(n-1)}{2}\,;\nonumber\\
&&d_{n,3}=\frac{n(n-1)(n-2)(3n-5)}{24}\,;\nonumber\\
&&d_{n,4}=\frac{n(n-1)(n-2)^2(n-3)^2}{48}\,.
\eea
It is clear from discussions in previous sections that
knowing the residue of the rational function $S_3(z)$
at $z=z_2$ (i.e. $w_{-2}^{(3,2)}$) is sufficient to fully recover the function
itself. Here is the final result for this residue
\bea
w_{-2}^{(3,2)}=\frac{A}{(z_2-1)^2}+\frac{B}{z_2^2}+\frac{C}{z_2(z_2-1)}\,,
\label{S3_functon}
\eea
where
\bea
&&A=\frac{2-n}{n^2}\,\, c_{0,1} \left(c_{0,1}+c_{2,1}\right) \left(n-c_{2,1}\right)\,;\nonumber\\
&&B=c_{1,3}-c_{2,3}+(n-2) \nonumber\\
&&\quad\times\left(\frac{1-n}{6n^2}\, c_{2,1}
\left(12 c_{2,1}^2-9 n c_{2,1}+2 n^2\right)
+c_{1,2} \left(1-\frac{2 c_{2,1}}{n}\right)
-c_{2,2} \left(1-\frac{3 c_{2,1}}{n}\right)\right)\,;\nonumber\\
&&C=2 \left(1-\frac{c_{2,1}}{n}\right)
\left(\left(\frac{2}{n}-1\right) c_{2,1} c_{0,1}
-c_{0,1}+c_{0,2}-c_{1,2}\right) \nonumber\\
&&\qquad+c_{0,1} \left(n-1+\frac{2}{n}\,\, \left(c_{1,2}
-c_{2,2}\right)\right)\,.
\label{S3_functon_ABC}
\eea
Recall now that the parameters $c_{0,1},c_{0,2}$
($c_{2,1},c_{2,2}$) are related to the fundamental
(anti-fundamental) hyper-multiplet masses $m_u$,
($\bar{m}_u$) as
\bea
&&c_{2,1}=\sum_{u=1}^nm_u;\qquad
c_{2,2}=\sum_{1\le u<v\le n}m_um_v\,;\nonumber\\
&&c_{0,1}=\sum_{u=1}^n\bar{m}_u;\qquad
c_{0,2}=\sum_{1\le u<v\le n}\bar{m}_u\bar{m}_v\,.
\eea
Due to eq. (\ref{trphiJ_vs_c})
\bea
c_{1,2}=-\frac{1}{2}\langle\tr \phi^2\rangle;
\qquad c_{1,3}=\frac{1}{3}\langle\tr \phi^3\rangle\,.
\eea
Thus (\ref{S3_functon}), (\ref{S3_functon_ABC}) go
beyond the standard AGT correspondance explicitly
relating the semiclassical four-point function including
a descendant field $W^{(3)}_{-2}V(z_2)$ with the gauge theory
expectation values $\langle\tr \phi^2\rangle$ and
$\langle\tr \phi^3\rangle$. The latter quantities can
be calculated either with direct instanton calculus or
DSW curve methods thus leading to the evaluation of
the Toda descendant including four point conformal block
\footnote{Notice that this conformal block can {\bf not}
be related to the respective block of primaries via Ward
identities, so that a straightforward CFT calculation is not
available.}.

%%%%%%%%%%%%%%%%%%%%%%%%%%%%%%%%%%%%%%%%%%%%%%%%%%%%%%%%%%%%%%%%%%%
\newpage

\section*{Conclusion}
As a final remark note that from the mathematical point of view
we have shown that:
\begin{itemize}
\item{
Any $n$'th order Fuchsian differential equation with generic ordinary singular points at $0$ and $\infty$,
additional $r+1$ ordinary points at $z_1=1,z_2,\ldots,z_{r+1}$
each of them obeying a set of indices of the form $\{0,1,\ldots,n-2,\xi_\alpha\}$ ($\xi_{\alpha}$ are arbitrary), such that the equation
doesn't accept any logarithmic solution can be represented in the form (\ref{diff_eq}), where $\chi_\alpha (x)$ are $n$'th order polynomials
with highest coefficient given by (\ref{chi_0}), (\ref{q_z_map}).}
\item{The monodromy problem of such differential equation is
intimately related to the system of difference equations (\ref{chi_alpha})
in the sense that the eigenvalues of the monodromy matrices
$M_\gamma \in SL(2,\mathbb{C})$ are given in terms of the period
integrals of the meromorphic differentials $xd\log(y_\alpha(x))$, where
$y_\alpha(x)$} are the solutions of difference equations (\ref{chi_alpha}).
\end{itemize}

%%%%%%%%%%%%%%%%%%%%%%%%%%%%%%%%%%%%%%%%%%%%%%%%%%%%%%%%%%%%%%%%%%%%%%%%
%%%%%%%%%%%%%%%%%%%%%%%%%%%%%%%%%%%%%%%%%%%%%%%%%%%%%%%%%%%%%%%%%%%
\vspace{1cm}

\centerline{\large\bf Acknowledgments}

\vspace{0.5cm}
I thank F.~Fucito, J.~F.~Morales and G.~Sarkissian
for many interesting discussions.
I am grateful to S.~Theisen for hospitality at the AEI Potsdam-Golm
where the bulk of this work has been completed and for a discussion.
A discussion with S.~Fredenhagen is also gratefully acknowledged.
This work was supported by the Armenian State Committee of Science
in the framework of the research project 15T-1C308 and by the
Volkswagen Foundation of Germany.

%%%%%%%%%%%%%%%%%%%%%%%%%%%%%%%%%%%%%%%%%%%%%%%%%%%%%%%%%%%%%%%%%%%%%%%%

%\bibliography{references_DSWFT}
\bibliographystyle{JHEP}

\providecommand{\href}[2]{#2}\begingroup\raggedright
\endgroup

\end{document}